\documentclass[12pt]{article}
\usepackage{graphicx}

\newcommand\ion[2]{#1$\;${\small\rmfamily\@Roman{#2}}\relax}%

\newcommand{\Msun}{${\rm M}_\odot$}

\newcommand{\kms}{km s$^{-1}$}

\def\eps@scaling{1.0}%

\newcommand\plottwo[2]{%
 \centering
 \leavevmode
 \columnwidth=.45\columnwidth
 \includegraphics[width={\eps@scaling\columnwidth}]{#1}%
 \hfil
 \includegraphics[width={\eps@scaling\columnwidth}]{#2}%
}%

\newcommand{\sovast}{Soviet~Ast.}
\newcommand{\mnras}{MNRAS}
\newcommand{\apj}{ApJ}
\newcommand{\apjl}{ApJ}
\newcommand{\nat}{Nature}
\newcommand{\apjs}{ApJS}
\newcommand{\aap}{A\&A}
\newcommand{\araa}{ARA\&A}
\newcommand{\pasp}{PASP} 
\newcommand{\aj}{AJ}

\usepackage{scicite}

\usepackage{times}

\newenvironment{sciabstract}{%
\begin{quote} \bf}
{\end{quote}}

\newcounter{lastnote}

\title{PTF~11kx: A Type-Ia Supernova with a Symbiotic Nova Progenitor}

\author
{ B.~Dilday$^{1,2\ast}$, D.~A.~Howell$^{1,2}$, S.~B.~Cenko$^{3}$,
J.~M.~Silverman$^{3}$, P.~E.~Nugent$^{3,4}$, \\
M.~Sullivan$^{5}$, S.~Ben-Ami$^{6}$,  L.~Bildsten$^{2,7}$, M.~Bolte$^{8}$, 
M.~Endl$^{9}$, 
A.~V.~Filippenko$^{3}$, \\
O.~Gnat$^{10}$, 
A.~Horesh$^{12}$, E.~Hsiao$^{4,11}$, M.~M.~Kasliwal$^{12, 13}$,
D.~Kirkman$^{14}$, \\
K.~Maguire$^{5}$, 
G.~W.~Marcy$^{3}$, 
K.~Moore$^{2}$, Y.~Pan$^{5}$, J.~T.~Parrent$^{1,15}$, \\
P.~Podsiadlowski$^{5}$, 
R.~M.~Quimby$^{16}$, 
A.~Sternberg$^{17}$, N.~Suzuki$^{4}$, D.~R.~Tytler$^{14}$, \\
D.~Xu$^{6}$, 
J.~S.~Bloom$^{3}$, 
A.~Gal-Yam$^{18}$, I.~M.~Hook$^{5}$, \\
S.~R.~Kulkarni$^{12}$, 
N.~M.~Law$^{19}$, 
E.~O.~Ofek$^{18}$, D.~Polishook$^{20}$, D.~Poznanski$^{21}$
\\
\\
 \scriptsize{$^{1}$Las Cumbres Observatory Global Telescope Network, 6740 Cortona Dr., Suite 102, Goleta, California 93117, USA}\\
 \scriptsize{$^{2}$Department of Physics, University of California, Santa Barbara,
 Broida Hall, Mail Code 9530, Santa Barbara, California 93106-9530, USA}\\
\scriptsize{$^{3}$Department of Astronomy, University of California, Berkeley,
 CA 94720-3411, USA}\\
\scriptsize{$^{4}$Lawrence Berkeley National Laboratory, Mail Stop 50B-4206, 1
 Cyclotron Road, Berkeley, California 94720, USA}\\
\scriptsize{$^{5}$Department of Physics (Astrophysics), University of Oxford, Keble
 Road, Oxford, OX1 3RH, UK}\\
\scriptsize{$^{6}$Department of Particle Physics and Astrophysics, The Weizmann Institute of Science, Rehovot 76100, Israel}\\
\scriptsize{$^{7}$Kavli Institute for Theoretical Physics, University of California, Santa Barbara, CA, 93106, USA}\\
\scriptsize{$^{8}$UCO/Lick Observatory, University of California, Santa Cruz, California 95064, USA}\\
\scriptsize{$^{9}$McDonald Observatory, The University of Texas at Austin, Austin, TX 78712, USA}\\
\scriptsize{$^{10}$Racah Institute of Physics, The Hebrew University of Jerusalem, 91904, Israel}\\
\scriptsize{$^{11}$Carnegie Institution of Washington, Las Campanas Observatory, Colina El Pino, Casilla 601, Chile}\\
\scriptsize{$^{12}$Cahill Center for Astrophysics, California Institute of Technology, Pasadena, CA, 91125, USA}\\
\scriptsize{$^{13}$Observatories of the Carnegie Institution for Science, 813 Santa Barbara St, Pasadena CA 91101 USA}\\
\scriptsize{$^{14}$Center for Astrophysics and Space Sciences, University of California San Diego, La Jolla, CA 92093-0424, USA}\\
\scriptsize{$^{15}$Department of Physics and Astronomy, Dartmouth College, Hanover, NH, USA}\\
\scriptsize{$^{16}$IPMU, University of Tokyo, 5-1-5 Kashiwanoha,
Kashiwa-shi, Chiba, 277-8583, Japan}\\
\scriptsize{$^{17}$Minerva Fellow, Max Planck Institute for
Astrophysics, Karl Schwarzschild St. 1, D-85741 Garching, Germany}\\
\scriptsize{$^{18}$Benoziyo Center for Astrophysics, The Weizmann
Institute of Science, Rehovot 76100, Israel}\\
\scriptsize{$^{19}$University of Toronto, 50 St. George Street, Toronto M5S 3H4, Ontario, Canada}\\
\scriptsize{$^{20}$Department of Earth, Atmospheric, and Planetary Sciences, Massachusetts Institute of Technology, Cambridge, MA 02139, USA}\\
\scriptsize{$^{21}$School of Physics and Astronomy, Tel-Aviv University, Tel-Aviv 69978, Israel}\\
\scriptsize{$^\ast$To whom correspondence should be addressed. E-mail: bdilday@lcogt.net}
}

\date{}

\begin{document} 


\baselineskip24pt


\maketitle 

\newpage

\begin{sciabstract}
  There is a consensus that Type-Ia supernovae (SNe~Ia) 
 arise from the thermonuclear explosion of
  white dwarf stars that accrete matter from a binary
  companion.  However, direct observation of SN~Ia
  progenitors is lacking, and the precise nature of the binary
  companion remains uncertain.  
  A temporal series of high-resolution optical spectra of the 
  SN~Ia PTF~11kx reveals a complex
  circumstellar environment that provides an unprecedentedly detailed view of  
  the progenitor system.
  Multiple shells of circumsteller are detected and the 
  SN ejecta are seen to interact with circumstellar material (CSM)
  starting $59$ days after the explosion.
  These features are best described by a symbiotic nova
  progenitor, similar to RS Ophiuchi.
\end{sciabstract}


Supernova PTF~11kx was 
discovered 
January 16, 2011 (UT) by the Palomar 
Transient Factory (PTF), at a 
cosmological redshift 
of 
$z = 0.04660 \pm 0.00001$\cite{DR8}.
This corresponds to a 
luminosity distance of $\sim 207$ Mpc.
The initial spectrum of PTF~11kx, taken January 26, 2011,
showed saturated absorption in the Ca~II
H\&K lines, along with weak absorption in the Na~I~D lines (Fig.~1).  
In interstellar gas, 
the Na~I~D lines are usually of strength comparable to (or greater than)
that of the Ca~II lines \cite{Welty_96};
thus, this combination of 
absorption features 
indicated that PTF~11kx may have affected its 
surroundings,
revealing
evidence of its
circumstellar environment and 
progenitor system.  
SNe~Ia are known to exhibit photometric \cite{Pskovskii_77, Phillips_93}
 and spectroscopic \cite{Nugent_95} diversity
that is correlated with intrinsic brightness, 
such that bright (faint) SNe~Ia have relatively broad (narrow) 
light curves, high (low) photospheric temperature, and weak (strong)
Si~II $\lambda$6150 absorption. 
Aside from the saturated Ca~II absorption,
the initial spectrum and a subsequent temporal series of spectra 
show PTF~11kx to resemble SN~1999aa
\cite{Li_01, Garavini_04}, a broad/bright SN~Ia (Fig.~1).
This suggests that insights into the progenitor of PTF~11kx may be
applicable to SNe~Ia generally, rather than to only 
a subset of peculiar objects. 
A complete sample of nearby SNe~Ia shows that the subclass 
similar to SN~1999aa comprises 
$\geq 9\%$ of all SNe~Ia \cite{Li_11a}.

We obtained high-resolution spectra ($R \approx 48,000$) 
of PTF~11kx
with the Keck~I High Resolution Echelle Spectrometer (HIRES) 
instrument at $-1$, $+9$, $+20$, and $+44$ days, relative
to $B$-band maximum light.  Low-resolution spectra were obtained on 14 epochs
between $-3$ and $+130$ days and show the composition and evolution of
features that are unique to SNe~Ia (Fig.~1).
The SN system is blueshifted from the galaxy cosmological 
redshift by $\sim 100$ \kms,
as determined from the strongest narrow interstellar Na~D absorption
line \cite{Patat_07, Sternberg_11}. 
An assumption that emission lines from a nearby H~II region evident in the
two-dimensional spectra are indicative of the progenitor velocity 
gives a consistent result.

The high-resolution spectra reveal narrow absorption lines of Na~I,
Fe~II, Ti~II, and He~I (Fig. 2).  
All are 
blueshifted by a velocity 
of $\sim 65$ \kms\ relative to the progenitor system, 
with a velocity dispersion of $\sim 10$ \kms, and are
consistent with arising in the same cloud or shell.  The sodium lines
increase in depth over time, a feature which has been resolved in
high-resolution spectra of two previous
SNe~Ia (SN 2006X and SN 2007le), 
and has been explained as the photoionization of CSM by the SN
light, followed by recombination that is detected through the
increased presence of
neutral sodium\cite{Patat_07, Simon_09}.  
A statistical study using high-resolution spectra of $35$ SNe Ia 
has shown that at least $20 \%$ of those with spiral host galaxies
have circumstellar material revealed
through narrow Na~I~D absorption lines \cite{Sternberg_11}.

However, narrow lines of circumstellar
Fe~II, Ti~II, and He~I have not been seen previously in a SN~Ia.
He~I $\lambda$5876 strengthens over time, which can be explained by photoionization
and recombination, if sufficient far-ultraviolet (UV) photons are generated in the 
early phases of the SN.
SNe~Ia are expected to produce only weak emission in the 
far-UV, but interaction of the SN ejecta with an 
extended progenitor such as a red giant star
can produce an excess of X-ray and UV photons \cite{Kasen_10}.
Lower excitation
states of Fe~II $\lambda\lambda$4923, 5018, 5169 decrease in strength in time,
while higher excitation states such as Fe~II $\lambda$5316 remain constant.  
A possible explanation for this is that the excitation energy of
Fe~II $\lambda$5316 (3.153 eV) corresponds to the saturated Ca~II~K line, 
and thus the population of this level is unaffected by the SN light.
Because SNe~Ia are weak sources in the UV, the distance over which they
can ionize gas is limited, and hence the observed
photoionization indicates that the material is circumstellar
\cite{Patat_07, Simon_09}.  
An alternative explanation, as the projection effect of different
interstellar clouds, was proposed for 
the time-variable Na~I~D lines in SN~2006X \cite{Chugai_08}; however, 
interaction with the CSM in PTF~11kx demonstrates
conclusively that CSM is present.
There are other narrow sodium features at different relative velocities that
do not vary with time, which must originate at a larger distance 
and are most likely interstellar (Fig. 2).

The hydrogen Balmer series is clearly detected in
absorption at $\sim$ 65 \kms, and is 
consistent with being 
at the same velocity as 
the Fe~II, Ti~II, Na~I, and He~I lines (Fig.~2). The H$\alpha$ and H$\beta$ 
lines show narrow
P-Cygni profiles, which are characteristic of an expanding shell of
radiating material.  The presence of circumstellar hydrogen has long
been identified as one of the characteristics expected in the
single-degenerate SN~Ia progenitor model\cite{Iben_84, Branch_95}, but 
has only been detected in at most two other cases, 
SN~2002ic \cite{Hamuy_03} and SN~2005gj \cite{Aldering_06}, 
and only at levels much stronger than in the early epochs of PTF~11kx.

These narrow features of H, He, Fe, Ti, and Na
arise from material that is distinct 
from the stronger Ca
absorption, which, owing to its more blueshifted absorption minimum of
$\sim 100$ \kms, must be at a higher velocity.  In fact, 
at later epochs, emission lines of Ca and H emerge, marking
the onset of interaction between SN ejecta and CSM.
This transition of circumstellar lines from absorption to emission has
not previously been observed in a SN Ia, and shows definitively
the presence of CSM.
Narrow
absorption components can be seen atop the broader emission 
(FWHM $\approx 1000$ \kms; Fig. 3),
indicating that there are two regimes of material, with
the faster-moving H and Ca interior to the slower-moving
shell (Fig.~4.)  

The saturation of the Ca~II~K line near maximum light requires that no
part of the SN photosphere is uncovered by the absorbing material;
this constrains the distance of the absorbing cloud from the progenitor 
to be further than
the SN ejecta, and the size of the cloud to be as large as the SN photosphere.
Assuming typical values for the 
ejecta expansion velocity of the outer layers of the exploding 
white dwarf of 25,000 \kms, 
and mean photospheric velocity of 10,000 \kms,  
these constraints are $\sim 4.3 \times 10^{15}$ cm and
$\sim 1.7 \times 10^{15}$ cm, respectively (Fig. 4).
The column density of Ca~II in the near-maximum-light spectra 
is determined to be $\sim 5 \times 10^{18} ~\mathrm{cm}^{-2}$ \cite{SOM}. 
Assuming that the Ca is part of a spherical shell 
of material with uniform density
would imply a total mass in the CSM ($\sim 5.3$ \Msun) 
which is incompatible with the photometric behavior of the SN \cite{SOM}. 
Therefore, the material must be non-uniform.
A small range of viewing 
angles that intersect such a dense cloud of absorbing material may 
partly explain
the rare occurrence of variable absorption features in SNe~Ia.

The Ca~II~H\&K lines detected in lower resolution spectra
confirm the disappearance of the absorption over $\sim 40$ days
and the emergence of a broad emission feature in the spectrum taken $+39$
days after maximum (Fig.~1).
The broad emission feature initially exhibits a
narrower absorption component as well
(possibly because of an outer shell of CSM; Fig. 4), 
but this is no longer evident after day $+56$.  At day $+39$,
H$\alpha$ also begins to exhibit a broad emission component.  
Assuming a rise time of $\sim 20$ days and an ejecta velocity
of 25,000 \kms\ would imply that the location of the 
interaction is $\sim 10^{16}$ cm.
The temporal evolution of the Ca~II K and H$\alpha$ lines are shown in Fig. 3.

A model for the progenitor of PTF~11kx must thus explain the presence
of multiple components of CSM, fast-moving material
interior to slower moving material, velocities of the narrow
absorption components that are larger than typical red giant winds, and a
region evacuated of CSM, leading to a delay between explosion and the
emergence of broad components of Ca and H emission.   A SN~Ia occurring 
in a symbiotic nova system naturally provides an
explanation for all of these features.  
In this model, accretion onto a near-Chandrasekhar-mass white dwarf occurs 
through the wind from a red giant star. The wind also deposits CSM into the 
system which is concentrated in the orbital plane \cite{Walder_08}. 
Episodic thermonuclear
runaways on the surface of the white dwarf cause 
nova events, which eject a mass of $\sim 10^{-7}$~\Msun\ at 
velocities of thousands of \kms. 
This process results in expansion velocities of 
CSM at $\sim 50$--100 \kms\ concentrated in the orbital plane, 
as is observed in high-resolution spectra of RS
Ophiuchi \cite{Patat_11}.  
The delay in the interaction between SN ejecta and CSM, leading 
to the emergence of the
broad H$\alpha$ and Ca~II emission, is explained as being due to an
evacuated region around the SN caused by the previous nova event
\cite{WoodVasey_06} that will be present until the
red giant wind has had time to replenish the local CSM.

The system of features seen in the spectra of PTF~11kx 
is inconsistent with expectations for double-degenerate 
SN~Ia progenitors. 
Double-degenerate SN~Ia progenitors
first go through a common-envelope phase where the outer layers of the
secondary are stripped, resulting in large amounts of CSM.
However, it is generically expected that the envelope
material will have dissipated by the time of the SN~Ia
event\cite{Chugai_04}. 
Recent modeling of the white dwarf merger process
in double-degenerate SN models 
suggests that heating in the accretion
disk can drive a wind, depositing CSM into the
system; however, because of the previous common-envelope phase, the
presence of hydrogen cannot be accounted for\cite{Shen_12}.
Alternative models for SN~2002ic and SN~2005gj which suggested that
they were not SNe Ia are also implausible for
PTF~11kx \cite{SOM}. 

For epochs less than $\sim 20$ days after maximum, 
the light curve of PTF~11kx resembles that of typical broad/bright
SNe-Ia \cite{SOM}.
Using the color law of the SiFTO 
\cite{Conley_08} light-curve model, we estimate that PTF~11kx is only
moderately extinguished by dust ($\sim 0.5$ mag in
$V$), and that the absolute magnitude ($M_{V} \approx -19.3$) is
as expected for a SN~1999aa-like SN~Ia. 
At epochs later than $\sim 20$ days, the light curve 
appears brighter than the SN~Ia templates, indicating that 
some brightening due to CSM
interaction has begun. At an epoch of 
$\sim +280$ days the SN is at an absolute magnitude of
$\sim -16.6$, roughly 3 mag brighter than expected,
indicating that circumstellar interaction is still ongoing.
Circumstellar interaction, including broad H emission and 
excess luminosity, was observed previously in SN~2002ic and SN~2005gj,
which had the appearance of the extreme broad/bright 
SN~1991T, superposed with features from strong CSM interaction.
The strength of the interaction casts some doubt on their
identity as SNe~Ia, which is not the case for PTF~11kx (Fig.~1).
Indeed, PTF~11kx bridges the observational gap between broad/bright SNe~Ia and
SN~2002ic/SN~2005gj, showing that SNe~Ia exist with 
CSM interaction that is weaker and
begins much later after explosion. 
This supports and enhances the interpretation 
that the earlier SNe with H
emission were SNe~Ia, where the strength of the CSM interaction
depends on details of the progenitor system, 
such as the mass-loss and accretion rates and the time since the last nova
eruption.

Measurements of the host-galaxy properties for SNe~Ia
suggest that broad/bright SNe~Ia 
arise from relatively young stellar populations\cite{Sullivan_06}, and 
a complete understanding of SN~Ia progenitors must account
for the occurrence of CSM interaction preferentially in young 
progenitor systems \cite{Sternberg_11}.
The host galaxy of PTF~11kx is a 
spiral galaxy with active star formation, and therefore has a 
substantial
fraction of young stars.
In contrast to SN~2002ic/SN~2005gj, the host galaxy has typical values
for mass and metallicity, so there is no evidence that PTF~11kx
must have come from an atypical stellar population.

There are several indications that white dwarf mergers are 
required to explain some SNe~Ia, including the rate in old stellar 
populations \cite{Maoz_10} and the
existence of SNe~Ia radiating at a luminosity that requires an exploding white
dwarf above the Chandrasekhar limit\cite{Howell_06, Silverman_11}.
The lack of a surviving companion near the 
remnant of a 
SN~Ia in the Large Magellanic Cloud has provided evidence 
for a double-degenerate progenitor for that particular case \cite{Schaefer_12}.
A symbiotic nova progenitor has been ruled out for the 
nearby SN~Ia PTF11kly/SN~2011fe \cite{Li_11},
which implies either a main sequence companion or 
a double-degenerate progenitor \cite{Nugent_11, Horesh_12}.
On the other hand, the detection of CSM, 
inferred through either
temporal variability \cite{Patat_07,Simon_09}
or statistical analysis
of the velocity \cite{Sternberg_11} 
of narrow absorption features, has provided recent evidence 
in support of the single-degenerate progenitor scenario for some SNe~Ia.
The remnant of Kepler's SN shows evidence for interaction with a
circumstellar shell which supports a single-degenerate progenitor for 
that SN~Ia \cite{Chiotellis_12}.

The system of features observed in 
PTF~11kx supplies direct evidence for a single-degenerate 
progenitor with a red giant companion, and
our data suggest a link between SNe~Ia that show weaker signatures of
the circumstellar environment (SN 2006X, SN 2007le, Kepler's SN) and
those that show stronger signatures (SN 2002ic, SN 2005gj). 
The fraction of SNe~Ia that exhibit prominent circumstellar 
interaction near maximum light is $\sim 0.1$--1\%, but
more subtle indications of a symbiotic progenitor
could be missed in typical SN observations \cite{SOM}.
Additionally, even more extreme interaction of a SN~Ia with CSM 
would not be distinguishable from a Type IIn SN, and it has been
speculated that 
some SNe~IIn are in fact SNe~Ia hidden within very strong CSM
interaction \cite{Hamuy_03}. 
Our data imply that there is indeed a wide range of CSM interaction
possible from SNe~Ia, and support this interpretation. If this is the
case, then current estimates of the rate of SNe with progenitors 
similar to that of PTF~11kx will be underestimated.
Theoretical expectations for the fraction of SNe~Ia from the 
symbiotic progenitor channel are 
$\sim 1$--30\% \cite{Han_04, Lu_09}, consistent with
observational constraints.

The solution to the long-standing puzzle of the progenitors of SNe~Ia 
is that there is no single progenitor path.
It remains to be determined 
in what proportion 
different progenitor channels contribute to the total rate of 
SNe~Ia, whether different channels result in 
different absolute luminosities or color evolution, 
and how the potential change of the relative contributions of 
different channels with look-back time may affect the 
use of SNe~Ia in accurately measuring cosmological expansion.




\bibliographystyle{Science}

\begin{enumerate}

\item[36.]\label{Law_09}
N.~M. {Law}, {\it et~al.\/}, {\it \pasp\/} {\bf 121}, 1395 (2009).

\item[37.]\label{Rest05}
A.~{Rest}, {\it et~al.\/}, {\it \apj\/} {\bf 634}, 1103 (2005).

\item[38.]\label{Miknaitis07}
G.~{Miknaitis}, {\it et~al.\/}, {\it \apj\/} {\bf 666} (2007).

\item[39.]\label{Becker_04}
A.~C. {Becker}, {\it et~al.\/}, {\it \apj\/} {\bf 611}, 418 (2004).

\item[40.]\label{Schecter_93}
P.~L. {Schechter}, M.~{Mateo}, A.~{Saha}, {\it \pasp\/} {\bf 105}, 1342 (1993).

\item[41.]\label{DR8}
H.~{Aihara}, {\it et~al.\/}, {\it \apjs\/} {\bf 193}, 29 (2011).

\item[42.]\label{Conley_08}
A.~{Conley}, {\it et~al.\/}, {\it \apj\/} {\bf 681}, 482 (2008).

\item[43.]\label{Turatto_03}
M.~{Turatto}, S.~{Benetti}, E.~{Cappellaro}, {\it From Twilight to Highlight:
  The Physics of Supernovae\/}, {W.~Hillebrandt \& B.~Leibundgut}, ed. (2003),
  p. 200.

\item[44.]\label{Poznanski_11}
D.~{Poznanski}, M.~{Ganeshalingam}, J.~M. {Silverman}, A.~V. {Filippenko}, {\it
  \mnras\/} {\bf 415}, L81 (2011).

\item[45.]\label{Jha_07}
S.~{Jha}, A.~G. {Riess}, R.~P. {Kirshner}, {\it \apj\/} {\bf 659}, 122 (2007).

\item[46.]\label{Holtzman_08}
J.~A. {Holtzman}, {\it et~al.\/}, {\it \aj\/} {\bf 136}, 2306 (2008).

\item[47.]\label{Panagia_06}
N.~{Panagia}, {\it et~al.\/}, {\it \apj\/} {\bf 646}, 369 (2006).

\item[48.]\label{Patat_11}
F.~{Patat}, {\it et~al.\/}, {\it \aap\/} {\bf 530}, A63+ (2011).

\item[49.]\label{Tremonti_04}
C.~A. {Tremonti}, {\it et~al.\/}, {\it \apj\/} {\bf 613}, 898 (2004).

\item[50.]\label{Iben_83}
I.~{Iben}, Jr., A.~{Renzini}, {\it \araa\/} {\bf 21}, 271 (1983).

\item[51.]\label{Trundle_08}
C.~{Trundle}, R.~{Kotak}, J.~S. {Vink}, W.~P.~S. {Meikle}, {\it \aap\/} {\bf 483}, L47 (2008).

\item[52.]\label{Benetti_06}
{Benetti}, S., {\it et~al.\/}, \apjl, {\bf 653}, L129 (2006).

\item[53.]\label{Livio_03a}
{Livio}, M. \& {Riess}, A.~G., \apjl, {\bf 594}, L93 (2003).

\item[54.]\label{Han_06}
{Han}, Z. \& {Podsiadlowski}, P., \mnras, {\bf 368}, 1095 (2006).

\item[55.]\label{Dilday_10}
{Dilday}, B.,  {\it et~al.\/}, \apj, {\bf 713}, 1026 (2010).

\item[56.]\label{Eck_02}
{Eck}, C.~R., {Cowan}, J.~J., \& {Branch}, D., \apj, {\bf 573}, 306 (2002).

\item[57.]\label{Vogt_94}
S.~S. {Vogt}, {\it et~al.\/}, {\it Society of Photo-Optical Instrumentation
  Engineers (SPIE) Conference Series\/}, {D.~L.~Crawford \& E.~R.~Craine}, ed.
  (1994), vol. 2198 of {\it Society of Photo-Optical Instrumentation Engineers
  (SPIE) Conference Series\/}, p. 362.

\item[]
  Some of the data presented herein were obtained at the W. M. Keck
  Observatory, which is operated as a scientific partnership among the
  California Institute of Technology, the University of California, and
  NASA; the observatory was made possible by the generous financial
  support of the W. M. Keck Foundation.
  The authors wish to recognize and acknowledge the very
  significant cultural role and reverence that the summit of Mauna Kea
  has always had within the indigenous Hawaiian community.  We are
  most fortunate to have the opportunity to conduct observations from
  this mountain. 
  We are grateful to the staffs of the Lick, Keck, and other observatories 
  for their assistance.
  We thank D.~Kasen for insightful discussions on the analysis
  presented here and we thank M.~Auger for assistance in using his 
  spectroscopic reduction pipeline. 
  The research of A.V.F.'s group was supported by
  grants from the NSF, the TABASGO Foundation, Gary \& Cynthia
  Bengier, and the Richard \& Rhoda Goldman Fund.
  M.M.K. acknowledges generous support from the Hubble Fellowship and
  Carnegie-Princeton Fellowship.
  A.G.Y. is supported by grants from the ISF and BSF, an
  ARCHES award, and the Lord Seiff of Brimpton Fund.
  The data presented in this paper are available from the 
  Weizmann Interactive Supernova data REPository 
  (http://www.weizmann.ac.il/astrophysics/wiserep/).

\end{enumerate}


\newpage 

\begin{figure}[t]
\includegraphics[width=5.75in, angle=0]{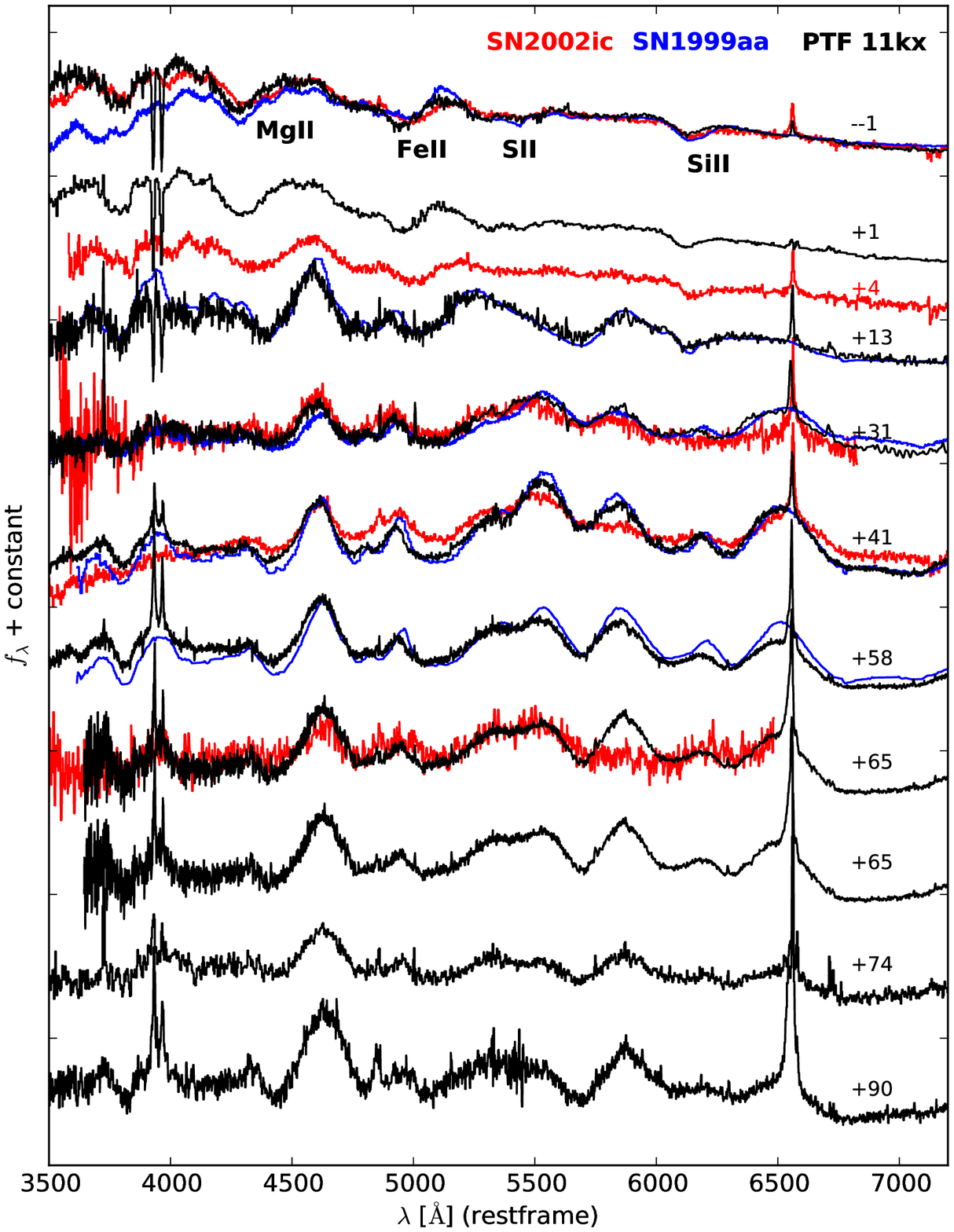}
  \caption{{\bf Comparison of spectra.}
    A temporal series of spectra of PTF~11kx is
    compared with that of 
    the broad/bright SN 1999aa \cite{Garavini_04}, 
    and the CSM-interaction SN~Ia, SN 2002ic \cite{Hamuy_03}, all at 
    a similar phase. 
    The location of Mg~II, Fe~II, S~II and Si~II present in the SN
    ejecta are labeled. 
The presence of both
    Si II and S II is a feature seen uniquely in SNe of Type Ia.
  }
\label{fig:sn99speccomp}
\end{figure}

\begin{figure}
\includegraphics[width=5.75in]{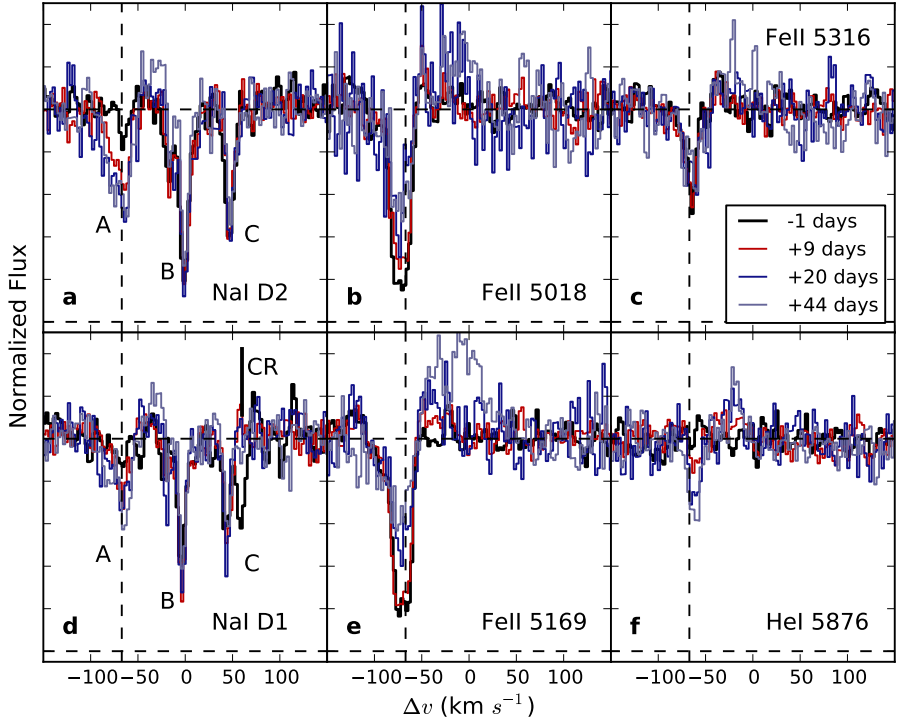}
  \caption{{\bf The temporal evolution of narrow absorption features in
    PTF~11kx.}  
 The panels show (a) Na~I~D2 , (b) Fe~II $\lambda$5018, (c)
    Fe~II $\lambda$5316, (d) Na~I~D1, (e) Fe~II $\lambda$5169, and (f) He~I 
    $\lambda$5876.
    Separate components of Na~I~D are labelled as A, B, and C.
    Component A strengthens with time and is interpreted as ionization
    from the SN followed by recombination, and thus is from
    circumstellar material.  Components B and C are attributed to the
    interstellar medium. The velocities are given relative to component B, 
    which is
    taken as indicative of the local velocity of the SN progenitor
    system.  The lower excitation states of Fe~II (panels (b) and (e))
    are seen to vary with time, while the higher excitation Fe~II line
    (panel (c)) appears to be constant in time. He~I absorption emerges in the
    later spectra.  
    The vertical dashed line marks the velocity of the circumstellar
    material at $-65$ \kms.  The horizontal dashed  
    lines mark 1 and 0 in normalized flux units.
    A cosmic ray in the day $-1$ spectrum
    near Na~I~D1 is marked by ``CR''; it produces a spurious
    absorption feature near the C component.  }
\label{fig:narrowlines}
\end{figure}

\begin{figure}[t]
\includegraphics[width=5.75in]{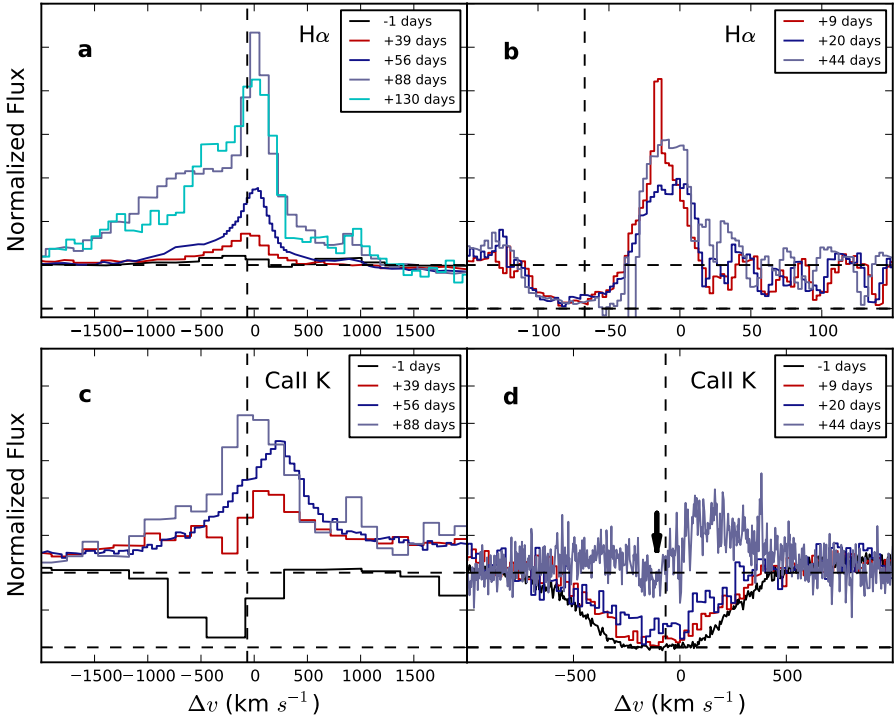}
  \caption{{\bf The temporal evolution of H$\alpha$ and Ca~II~K features
      in PTF~11kx.}  
    (a) Low-resolution H$\alpha$, 
    (b) high-resolution H$\alpha$,
    (c) low-resolution Ca~II~K, and 
    (d) high-resolution Ca~II~K. 
    The $+9$ and $+20$ day high-resolution Ca~II spectra 
    have been rebinned to $20$ \kms\ precision for clarity of presentation.
    The vertical dashed line marks the velocity of the
    slower circumstellar material at $-65$ \kms.  
    The large optical depth
    in the H$\alpha$ line causes an apparent blueshift in the 
    absorption minimum, relative to the $-65$ \kms\ expansion 
    velocity of the narrow lines shown in Fig. 2.
    The horizontal dashed lines mark $0$ and $1$ in normalized flux units
    and show that at early times the Ca~II lines are saturated.  
Narrow absorption, indicated by an arrow, 
    can be seen superposed on
    the broad emission in the day +44 spectrum in panel (d). Narrow 
    interstellar lines of Ca~II, corresponding to components B and C in 
    panels (a) and (d) 
    of Fig. 1, can be seen at $0$ \kms\ and $+45$ \kms. Several
    spectra in (a) and (c) appear to show narrow absorption superposed
    on broad emission, although the absorption is not always
    resolved.  
    This indicates that, 
    subsequent to the onset of CSM interaction,
    slower-moving material 
    external to the inner shell of interacting material remains,     
    possibly due to a decelerated nova shell.}
\label{fig:broadlines}
\end{figure}

\begin{figure}
\includegraphics[width=5.75in]{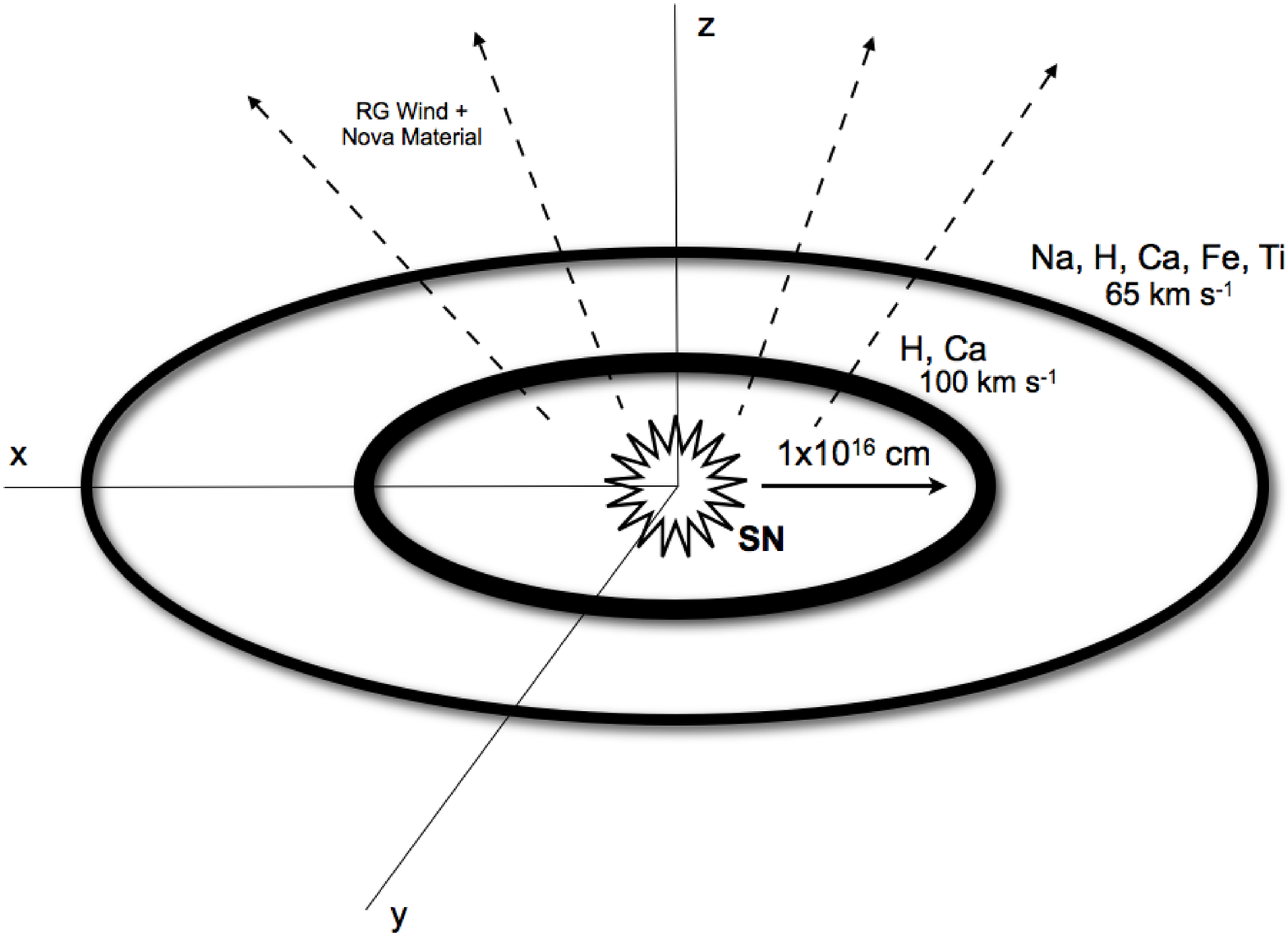}
  \caption{{\bf Schematic showing the interpretation of the
      observations of PTF~11kx.}  
    In the symbiotic nova progenitor system,
    a red giant wind deposits CSM into the 
    system which is concentrated in the orbital plane. 
    Episodic nova events further shape the CSM, 
    resulting in expansion velocities of $\sim 50$--100 \kms.
    There is an inner region of material 
    containing at least H and Ca, moving at 
    $\sim 100$ \kms, surrounded by more distant material containing H,
    Ca, Na, Fe, Ti, and He, moving at $\sim 65$ \kms.  
    The presence of hydrogen
    in the inner region is inferred from the onset of emission, concurrent
    with the onset of Ca~II emission. 
    This geometry, and the delay in the onset of CSM emission, is consistent
    with a relatively recent nova whose ejecta are sweeping up the
    circumstellar material and decelerating.  The inner boundary of CSM 
    is at a distance of $\sim 10^{16}\, \mathrm{cm}$,
    as determined from the onset of interaction.  
}
\label{fig:schematic}
\end{figure}




{\bf \large Supplementary Material}

\vspace{10mm}

{\bf S1. Spectroscopy}

\vspace{5mm}

Optical spectra of PTF~11kx were obtained with several
different telescopes: the Lick 3~m, Palomar 5~m, KPNO 4~m, WHT,
Keck, and Gemini North.
Spectra from the High Resolution Echelle Spectrometer (HIRES) 
on Keck~I were reduced using the 
Mauna Kea Echelle Extraction ({\tt makee}) 
software package\footnote{http://spider.ipac.caltech.edu/staff/tab/makee}. 
Spectra from other telescopes
and instruments were reduced using standard methods.
Journals of observations are given in Tables S1 and S2; UT
dates are used throughout this paper.
A time series of low-resolution spectra for PTF~11kx is shown in
Fig.~1.

\vspace{10mm}

{\bf S2. Photometry}

\vspace{5mm}

Photometry in the $r$ band was collected as part of the
normal PTF transient search on the Palomar 48~in telescope ({\it 36}).
PTF~11kx was also followed by the 
2~m Faulkes Telescope North. 
Between January 27, 2011 and June 9, 2011 (133 days), 
46 epochs were obtained in the SDSS $g$ and $r$ bands,
and 45 epochs in the SDSS $i$ band. Photometric observations 
of PTF 11kx were resumed on September 30, 2011.
At the location of PTF~11kx there is a substantial background 
due to the host galaxy. When this is the case, photometry 
for SNe is usually accomplished through difference imaging analysis,
where an image without the SN present is used as a difference imaging template.
Reference images for PTF~11kx have not yet been taken on the Faulkes North,
and so only preliminary photometry is currently available. 
To at least roughly subtract the host galaxy, we 
used SDSS images as difference imaging templates. 
Visual inspection of the difference images reveals that the host-galaxy
background is subtracted successfully, and thus
the use of templates from a 
different instrument
does not significantly 
affect the conclusions that PTF~11kx is a member of the 
broad/bright SN class. Initial photometry suggests that PTF~11kx
declines more slowly at epochs later than 
$\sim +20$ days than the templates of the same class, 
which is another indication of circumstellar interaction, and
late-time ($\sim 280$ days) photometry confirms this.
Data were processed using the {\tt photpipe} 
reduction pipeline ({\it 37, 38}). Difference imaging
was done using {\tt hotpants} ({\it 39}). 
Source detection and PSF photometry was done
using {\tt dophot} ({\it 40}). 
The photometry was calibrated by comparison of stars in the 
field of PTF~11kx to magnitudes provided by the SDSS DR8 
database ({\it 41}).

\vspace{10mm}

{\bf S3. SN Light Curve}

\vspace{5mm}

Comparing the colors of the SN to templates 
from the SiFTO SN light-curve model ({\it 42}), 
we estimate that the SN is only moderately extinguished by 
dust, with the extinction
amounting to $\sim 0.5$ mag in the $V$ band. 
It is commonly assumed that the equivalent width of the Na~D lines is
correlated with the value for extinction by dust (e.g.~{\it 43}),
and so the small value of the equivalent width of the non-variable
Na~D lines also supports this conclusion.
While this correlation has been shown to have significant 
scatter ({\it 44}), this is most relevant for
equivalent widths larger than those measured in PTF~11kx.
Assuming a concordance 
cosmology ($H_{0} = 70$ \kms~Mpc$^{-1}$), the absolute $B$-band magnitude of PTF~11kx is 
$\sim -19.3$, which is on the bright end of the 
SN~Ia luminosity function
and is consistent with the
observed distribution of high-stretch SNe~Ia.
The date of maximum
light (in the $B$ band) of PTF~11kx is January 29, 2011. 
Comparisons with the broad SN~Ia~1999aa ({\it 45}), 
the ``normal'' SN~Ia~2002er ({\it 45}),
and the CSM-interaction SNe~Ia~2002ic ({\it 17}) 
and 2005gj ({\it 46}), are shown in Fig.~S1.
While the photometry is preliminary, we can draw a few important conclusions:
(i) the light curve is of the high-stretch class and is 
qualitatively similar to that of SN~1999aa and SN~1991T,
(ii) the light curve is markedly dissimilar to that of SN~2002ic 
and SN~2005gj, and 
(iii) the absolute magnitude at peak is consistent with the population
of previously observed SNe~Ia.

\vspace{10mm}

{\bf S4. Radio Observation}

\vspace{5mm}

PTF~11kx was observed on March 30, 2011 in the X-band using the 
Expanded Very Large Array (EVLA), yielding a nondetection with a $1\sigma$
root-mean-square of 23 $\mu$Jy. Using a model of SN interaction with 
a smooth wind puts an upper limit on the inferred mass-loss 
rate in the wind ({\it 47}) of 
$\sim 4.5 \times 10^{-6}$ \Msun\ yr$^{-1}$ ($v_{\mathrm{w}}$/10 \kms)$^{1.65}$, 
where $v_{\mathrm{w}}$ is the velocity of the wind. 
The radio emission strength depends on microphysical parameters such as
the strength of the magnetic fields generated in the interaction 
(the $\epsilon_{B}$ parameter),
and if these magnetic fields are weaker than has generally been assumed,
then the constraints on the rate of mass loss 
would be less strict by perhaps an order of magnitude ({\it 31}). 
Estimates of the mass-loss rate for symbiotic novae such as 
RS Ophuichi are $\sim 3 \times 10^{-7}$ \Msun\, yr$^{-1}$, well
below our constraint ({\it 48}).

\vspace{10mm}

{\bf S4. Host-Galaxy Properties}

\vspace{5mm}

The host galaxy of PTF 11kx is a late-type, spiral galaxy. The
properties of the host galaxy, such as the star-formation rate, mass, 
and metallicity, are not unusual in any way. The absolute
magnitude is $M_{g} \approx -19.47$, and the metallicity has been derived
by ({\it 49}) to be 12 + log(O/H) $\approx 8.93$, which is slightly larger
than solar. This indicates
that PTF 11kx does not obviously arise from an atypical local
environment. As a comparison, the host galaxies for SN~2002ic
and SN~2005gj were low-luminosity galaxies with corresponding low
metallicities, which has been assumed to play a role in the
presence of CSM in the SN systems. PTF~11kx demonstrates that CSM
interaction can occur in a SN Ia even in a more normal galactic
environment. 

\vspace{10mm}

{\bf S5. Alternative Models for CSM-Interaction SNe Ia}

\vspace{5mm}

As discussed in the main text, PTF~11kx bears similarities to
SN~2002ic, 
an apparent SN~Ia with circumstellar interaction. Subsequent to
the discovery of SN~2002ic, several alternative models were proposed as
possible explanations. 
Given the association of PTF~11kx with SN~2002ic we question whether
any of those models are viable alternatives for PTF~11kx. 
In PTF~11kx, the multiple shells of CSM,
the nonuniform distribution of the CSM, and the long delay between
explosion and circumstellar interaction are important additional
constraints. \\

{\bf The ``SN 1.5'' Model:}
The ``SN 1.5'' model involves the explosion of the degenerate core of a
massive star ({\it 50}). 
This model cannot explain the long delay in the onset of
emission, the nonuniform distribution of CSM, or the multiple shells of
material seen in PTF 11kx. \\

{\bf A Luminous Blue Variable Progenitor:}
The arguments for an LBV progenitor applied only to SN
2005gj ({\it 51}), which had 
much stronger CSM interaction than SN 2002ic, and were
based mainly on the double P-Cygni profile in the H$\alpha$ line, 
a feature which
is not present in PTF 11kx. \\

{\bf A Peculiar SN Ic:}
The possibility that SN~2002ic was a SN Ic ({\it 52})
was based on the possible
presence of Mg and O in the late-time spectra.
The spectra of PTF~11kx shown in Fig.~1 definitively
rule out a core-collapse SN in this case. \\

{\bf A Double-Degenerate SN Ia:}
Subsequent to the suggestion that SN~2002ic had a double-degenerate
progenitor ({\it 53}), additional modeling has shown that it is not
expected that any significant circumstellar material from the 
common-envelope phase can remain at the time of the SN explosion ({\it 22}).
Moreover, modeling of the possible CSM due to wind generated during the
merger process of a double-degenerate progenitor cannot account for the
presence of hydrogen ({\it 23}). A double-degenerate progenitor 
also does not explain the multiple
distinct shells of CSM.  \\

{\bf Supersoft, Single-Degenerate Progenitor:}
The model of ({\it 54}) could possibly explain many of
the features of PTF 11kx, but our symbiotic nova model naturally
explains the low expansion velocity of the CSM (65 \kms~for PTF~11kx
instead of $\sim 100$ \kms), the multiple shells of
CSM, and the delay between explosion and interaction, while the 
({\it 54}) model does not. 

\vspace{10mm}

{\bf S6. Rate of SNe Ia with Symbiotic Nova Progenitors}

\vspace{5mm}

One of the features of PTF~11kx that supports the
conclusion that it is from a symbiotic nova progenitor is the prominent 
H and Ca emission seen $\sim 59$ days after explosion. 
The SDSS-II SN Survey had a high degree
of spectroscopic completeness for SNe~Ia at $z<0.15$, and the discovery of
one CSM-interaction SN~Ia (SN~2005gj) in a sample of 80 SNe~Ia with
$z<0.15$ implies a rate of such SNe of $\sim 1\%$ ({\it 55}). 
The Palomar Transient Factory currently has
discovered $\sim 1000$ SNe~Ia with one known example of a
CSM-interaction SN Ia (PTF~11kx), but the spectroscopic
completeness has not yet been determined or quantified and so this can
only be considered an order of magnitude estimate for the rate of
CSM-interaction SNe. 

However, an important conclusion from the discovery of PTF~11kx is that SNe~Ia
exist that show CSM-interaction at weaker levels and later onset
than SN~2002ic and SN~2005gj. This suggests that there is a continuum
of CSM-interaction in SNe Ia. If the continuum of interaction strength
extends to lesser values and later onsets, then the signs of interaction
would be missed in a large fraction of SNe Ia, as they do not typically
have high-signal-to-noise ratio spectroscopic observations at epochs $> 60$
days after explosion. The discovery of PTF~11kx highlights the importance 
of more extensive spectroscopic and photometric follow-up monitoring for
SNe~Ia extending to late times to determine the rate of SNe~Ia with
significant circumstellar material from the progenitor
system. 
Furthermore, CSM-interaction SNe~Ia with greater interaction
strength could be classified as SNe~IIn and these would contribute an
undetermined fraction to the overall SN~Ia rate.
Therefore, we can only say that SNe~Ia with prominent CSM interaction
occurring near maximum light is $\sim 0.1$--1\%. 
Population synthesis modeling predicts the fraction of SNe~Ia from the
symbiotic binary channel to be $\sim 1\%$ ({\it 33}) to as much as 
30\% ({\it 34}), which encompasses the range of observational 
constraints.

The existence of a red giant companion could in principle be
detected through radio emission due to interaction of the SN ejecta 
with the wind from the secondary, but no 
SN~Ia has yet been detected in the radio. However, the mass-loss limits
derived via radio nondetections ({\it 47, 56}) do not
generally constrain RS Oph-like systems which has a mass-loss rate of
$\sim 3 \times 10^{-7}$ \Msun\, yr$^{-1}$. Moreover,
theory and observations suggest that the CSM distribution in symbiotic
novae is nonuniform and the expected radio emission taking this into
consideration has not yet been modeled.
It is therefore unclear what effect this has on the limits derived from
nondetection, but it is plausible that this would introduce a viewing-angle dependence that would decrease the chance for detecting the radio signal.

\vspace{10mm}

{\bf S7. H$\alpha$ and Ca~II Fluxes}

\vspace{5mm}

Starting with the day $+39$ spectrum, a broad component of the 
H$\alpha$ emission begins to appear; 
the profile can be decomposed into broad and narrow components.
The full width at half-maximum intensity (FWHM) and the integrated 
fluxes in the broad component are given in Table S3. 
Starting with the day $+56$ spectrum, the Ca~II emission
begins to have a Gaussian appearance and the residual 
absorption seen in earlier spectra appears to be gone.
The FWHM and the integrated fluxes in the Ca~II H\&K lines are given in 
Table S4. To derive these values, the spectra were first 
spectrophotometrically calibrated to the $r$-band photometry from the
Faulkes 2~m telescope. The Gaussian components of the emission feature 
were then fit using the deblending function in the {\tt iraf} 
{\tt splot} procedure.

\vspace{10mm}

{\bf S8. Mass Estimates}

\vspace{5mm}

In the near-maximum-light spectra, the equivalent widths of the 
Ca~II H\&K lines are $\sim 10$~\AA. The lines are saturated, and 
on the square-root portion of the curve of growth, so that 
the equivalent width, $W_{\lambda}$, is given by
\begin{equation}
W_{\lambda} = (N \frac{\lambda_0^{4}}{2 \pi c} \frac{g_u}{g_l} A_{ul} \gamma_{u})^{1/2},
\end{equation}

\noindent where $N$ is the column density, 
$\lambda_0$ is the wavelength of the transition,
$A_{ul}$ is the Einstein spontaneous emission coefficient, 
$\gamma_{u}$ is the radiation damping constant, 
and $g_u$ and $g_l$ 
are the statistical weights of the upper and lower states, respectively. 
For the Ca~II~K line, with $\lambda_0 = 3933.6$~\AA, this results in 
$N = 4.44 \times 10^{16} ~W_{\lambda}^{2} ~\mathrm{cm}^{-2}$.
Thus, the column density in Ca~II is $\sim 5 \times 10^{18}~ \mathrm{cm}^{-2}$.
We write the total Ca mass as 
$M_{\mathrm{Ca}} = k \times 4 \pi r^{2}$, where $r$ is the radius at which the
material exists and $k$ is the covering fraction. The radius, $r$, 
can be estimated from the velocity of the SN ejecta, which 
we take to be $v \approx$ 25,000 \kms, and the time
at which the Ca goes into emission, which is $\sim +59$ days 
after explosion. This results in 

\begin{equation}
M_{\mathrm{Ca}} = 3.43 \times 10^{-4}\, k\, 
(\frac{v}{25,000 ~\mathrm{km}~\mathrm{s}^{-1}} 
\frac{t}{59 ~\mathrm{days}})^{2} 
(\frac{N}{5 \times 10^{18}~ \mathrm{cm}^{-2}}) 
 ~{\rm M}_\odot .
\end{equation}

Assuming a solar composition for the CSM, 
a value of $k=1$ would imply a total mass in the CSM shell of 
$\sim 5.3$ \Msun.
Modeling of the light curve for the CSM SNe 1997cy and 2002ic 
results in an estimate of the mass in the CSM for
those SNe of several solar masses ({\it 22}).
The total luminosity of PTF~11kx is much less, and the decline rate much
greater than that of either SN~1997cy or SN~2002ic, implying that the total CSM
mass is also much less.
Thus, we conclude that $k \ll 1$, and that the CSM material 
that generates the Ca~II absorption is not uniformly distributed.

\newpage

\makeatletter 
\renewcommand{\thefigure}{S1} 

\begin{figure}[t]
\begin{center}
\includegraphics[width=5.75in]{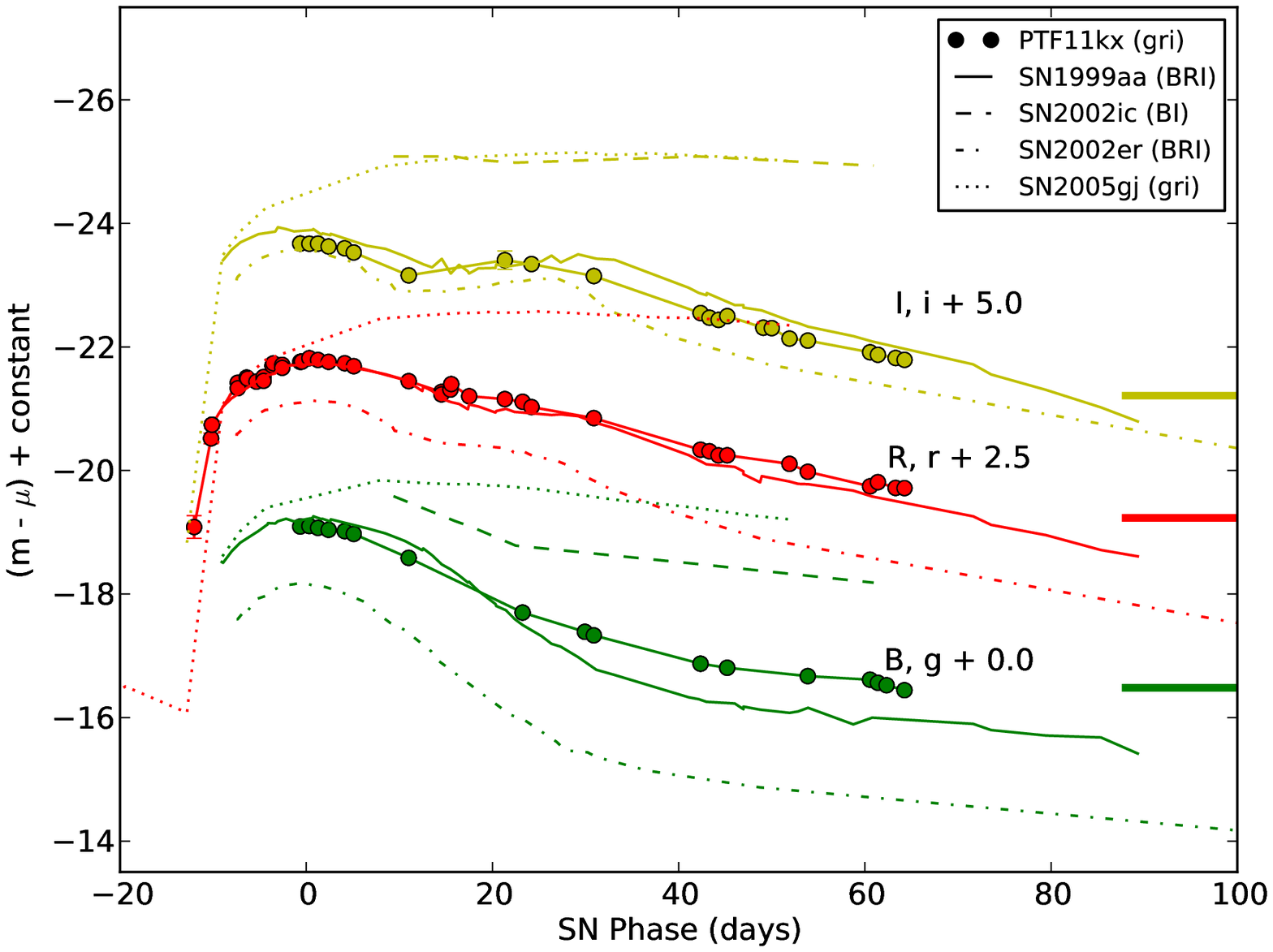}
\caption{
Comparison of the preliminary light curve of PTF~11kx to that of
other SNe~Ia. The data are the observer-frame magnitudes, without 
$K$-corrections. The PTF~11kx and SN 2005gj photometry is given in 
$gri$, and the other SNe in 
some combination of $BRI$. The red, green, and yellow points denote 
$B/g$, $R/r$, and $I/i$, respectively.
The broad/bright 
SN~1999aa (solid) 
and the ``normal'' 
SN~2002er (dash-dot) 
are shown for comparison. Also shown are 
the CSM-interaction 
SNe~Ia~2002ic (dashed) 
and 
2005gj (dotted).
PTF~11kx is much fainter than SN~2005gj and SN~2002ic, 
indicating a weaker/later onset
 of CSM interaction. The late-time, approximately constant, 
brightness of PTF~11kx is marked by the horizontal lines on the right
 side of the figure.
Spectroscopically and photometrically, PTF~11kx bridges the 
observational gap between SN~1991T/SN~1999aa and SN~2002ic/SN~2005gj.
}
\end{center}
\label{flc}
\end{figure}



\begin{table}[t]
\centerline{
\begin{tabular}{|l|l|c|r|r|r|}
\hline
UT Date & 
Exposure Time ($s$) & 
SN Phase (days) & 
Wavelength Range & 
Median S/N  & 
R $(\lambda/\Delta \lambda)$\\
\hline
20110128 & 1000 &  -1 & 3682-6538 & 10.4 & 48000 \\
20110207 & 1800 &  +9 & 4017-8564 &  6.2 & 48000 \\
20110218 & 1800 & +20 & 3682-7993 &  3.5 & 48000 \\
20110314 & 3600 & +44 & 3682-7992 &  4.2 & 48000 \\
\hline
\end{tabular}
}
\vspace{3mm}
Table S1. 
Journal of high-resolution spectroscopic 
observations. All spectra were taken with the HIRES instrument 
({\it 57})
on the Keck I telescope.
\end{table}

\begin{table}[t]
\centerline{
\begin{tabular}{|l|l|c|c|}
\hline
UT Date & 
Telescope & 
SN Phase (days) & 
Wavelength Range (\AA) \\
\hline
20110126 & Lick 3m          &                   -3 & $3448 - 10142$  \\
20110128 & Palomar 200in          &             -1 & $3490 -  9750$  \\
20110202 & Lick 3m          &                   +4 & $3440 -  9754$  \\
20110203 & KPNO 4m          &                   +5 & $3487 -  7570$  \\
20110209 & Lick 3m          &                  +11 & $3430 - 10138$  \\
20110221 & WHT          &                      +23 & $3190 -  9279$  \\
20110227 & Lick 3m          &                  +29 & $3400 - 10000$  \\
20110309 & Keck1          &                    +39 & $3330 - 10188$  \\
20110312 & Keck1          &                    +42 & $3948 -  6128$  \\
20110326 & Keck1          &                    +56 & $3125 - 10233$  \\
20110402 & Gemini North          &             +63 & $3815 -  9701$  \\
20110411 & KPNO 4m          &                  +72 & $3450 -  8451$  \\
20110427 & Keck1          &                    +88 & $3100 - 10200$  \\
20110608 & WHT          &                     +130 & $3036 -  9499$  \\
\hline
\end{tabular}
}
\vspace{3mm}
Table S2. 
Journal of low-resolution spectroscopic 
observations.
\end{table}

\begin{table}
\centerline{
\begin{tabular}{|l|l|c|c|}
\hline
UT Date & 
SN Phase (days) & 
$\sigma_v$ (\kms) &
Flux (ergs s$^{-1}$) \\
\hline
20110309  &   +39   & 600   &   1.64e+39   \\ 
20110326  &   +56   & 980   &   4.36e+39   \\ 
20110402  &   +63   & 1050   &   4.60e+39   \\ 
20110427  &   +88   & 1590   &   1.84e+40   \\ 
20110608  &   +130   & 2040   &   1.29e+40   \\ 
\hline
\end{tabular}
}
\vspace{3mm}
Table S3. 
H$\alpha$  velocity dispersion and integrated flux values.
\end{table}

\begin{table}[t]
\centerline{
\begin{tabular}{|c|c|c|c|c|c|}
\hline
\normalsize{UT Date} &
\normalsize{SN Phase} & 
\normalsize{CaII K $\sigma_v$} &
\normalsize{CaII K Flux} &
\normalsize{CaII H $\sigma_v$} &
\normalsize{CaII H Flux} \\
 &
(days) & 
(\kms) &
(ergs s$^{-1}$) &
(\kms) &
(ergs s$^{-1}$) \\
\hline
20110326  &   +56   & 780   &   4.56e+39  & 610   &   2.75e+39  \\ 
20110402  &   +63   & 680   &   2.56e+39  & 380   &   1.10e+39  \\ 
20110427  &   +88   & 850   &   4.42e+39  & 970   &   2.72e+39  \\ 
\hline
\end{tabular}
}
\vspace{3mm}
Table S4. Ca~II H \& K velocity dispersions and integrated 
flux values.
\end{table}

\end{document}